# Non-Hermitian funneling in anisotropic media

Yuan Tian[1, *], Nankun Gao[1, *], Xiujuan Zhang[1, †], Ming-Hui Lu[1, 2, 3, ‡], and Yan-Feng Chen[1, 3, §]

[1]*State Key Laboratory of Solid State Microstructures and Department of Materials Science and Engineering, Nanjing University, Nanjing 210093, China*
[2]*Jiangsu Key Laboratory of Artificial Functional Materials, Nanjing 210093, China*
[3]*Collaborative Innovation Center of Advanced Microstructures, Nanjing University, Nanjing 210093, China*

[*]*These authors contributed equally.*
[†‡§]*Email: xiujuanzhang@nju.edu.cn; luminghui@nju.edu.cn; yfchen@nju.edu.cn*

## Abstract

Non-Hermitian skin effect (NHSE) has emerged as a distinctive phenomenon enabling non-Bloch wave manipulation. However, it has been limited to discrete lattices requiring fine-tuned onsite gain/loss or asymmetric couplings. Here, moving beyond these discrete models, we realize novel NHSE in uniform media by leveraging anisotropy of non-Hermitian density tensors. Experiments based on an acoustic anisotropic metamaterial demonstrate that enabled by the NHSE, wave energy can be directed toward and collected at specific boundaries, exhibiting broadband and wide-angle characteristics. This intriguing phenomenon is termed non-Hermitian wave funneling, which, remarkably, occurs under uniform non-Hermitian modulations, free of fine-tuning. Furthermore, we identify a second-order NHSE, enabling wave funneling toward corners. Our work establishes a paradigm for exploring NHSE in uniform media, advancing the fundamental understanding of non-Hermitian physics and providing novel mechanisms for non-Bloch wave control in metamaterials or even natural materials without delicate tuning.

## Introduction

Quantum mechanics, traditionally built on Hermitian Hamiltonians for closed systems, faces challenges in describing real-world open systems that exchange energy with their environment. Non-Hermitian Hamiltonians resolve this by naturally accommodating gain and loss through complex eigenvalues and eigenstates [1,2], extending quantum frameworks to nonconservative regimes. This formalism reveals remarkable phenomena like parity-time (PT) phase transitions and exceptional points [3–6], enabling unidirectional wave propagation [7], coherent perfect absorption [8], and enhanced sensitivity [9–11].

Recently, non-Hermiticity has been brought forward to interplay with topological physics, leading to the discovery of non-Hermitian skin effect (NHSE) [12–20]. It describes under specific non-Hermitian modulation, a unique point-gap topology emerges, enabling non-Bloch wave control like unidirectional amplification and dynamical properties such as self-healing [21], self-acceleration [22] and edge burst [23,24]. While NHSE has been demonstrated in a variety of systems including photonics [25,26], quantum walks [27], phononics [28–31], mechanics [32–35] and electric circuits [36–39], these implementations are confined to tight-binding discrete lattices with fine-tuned asymmetric couplings or complex gain/loss modulations.

Here, we transcend discrete models by realizing NHSE in uniform media. Inspired by the theoretical links between anisotropy and non-Hermiticity [40,41], we design an acoustic metamaterial with subwavelength multilayers, which can be perceived as an anisotropic uniform medium. Integrated with a non-Hermitian sponge layer, this medium features an anisotropic complex density tensor that directly facilitates NHSE. Enabled by this NHSE, wave energy can be funneled along designed paths and localized at targeted boundaries across broadband frequencies and wide-angle incidences, as schematically illustrated in Fig. 1. By direct generalization, we can also realize second-order NHSE that funnels waves to corners. NHSE in uniform media represents a novel bulk-boundary relation beyond periodicity, opening a new playground for non-Hermitian physics. The minimal tuning requirements also make real applications highly feasible, not only in metamaterials well-known for their unprecedented material parameters, but also in natural materials where uniformity is readily accessible.

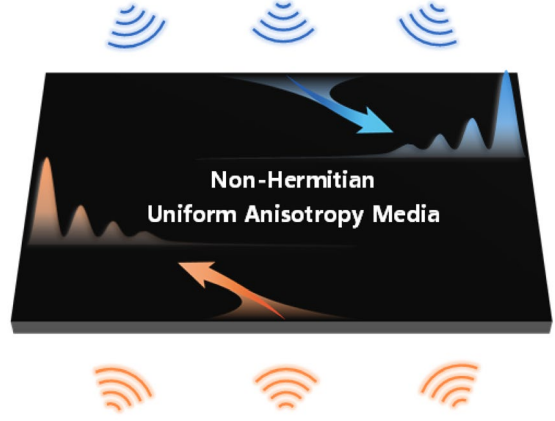

FIG. 1. Schematic of NHSE in anisotropic uniform medium, enabling non-Hermitian wave funneling along designed paths and localization at targeted boundaries, exhibiting broadband and wide-angle characteristics.

**Theoretical model**

We consider a two-dimensional (2D) uniform medium with an anisotropic density tensor $\rho = \begin{pmatrix} \rho_{xx} & \rho_{xy} \\ \rho_{yx} & \rho_{yy} \end{pmatrix}$ and bulk modulus $K$. Acoustic wave propagation is governed by the following wave equation [42]

$$\nabla \cdot (-\rho^{-1}\nabla p) - \frac{\omega^2 p}{K} = 0, \qquad (1)$$

and dispersion relation

$$\frac{\omega_{w/o}^2}{K} = \eta_{xx}(k_x + q_x k_y)^2 + \eta_{yy}(k_y + q_y k_x)^2 - (\eta_{xx} q_x^2 k_y^2 + \eta_{yy} q_y^2 k_x^2), \qquad (2)$$

where $p$ represents the pressure field, $\omega$ ($\omega_{w/o}$) the angular frequency, $k_x$ and $k_y$ the wave vectors along the *x*- and *y*-directions, respectively. $\rho^{-1}$ is replaced by a new notation $\eta$ defined as

$$\eta \equiv \rho^{-1} = \begin{pmatrix} \eta_{xx} & \eta_{xy} \\ \eta_{yx} & \eta_{yy} \end{pmatrix} = \frac{1}{\rho_{xx}\rho_{yy} - \rho_{xy}\rho_{yx}} \begin{pmatrix} \rho_{yy} & -\rho_{xy} \\ -\rho_{yx} & \rho_{xx} \end{pmatrix}, \qquad (3)$$

with material anisotropy characterized by $q_x$ and $q_y$, obeying

$$q_x = \frac{\eta_{xy} + \eta_{yx}}{4\eta_{xx}}, \quad q_y = \frac{\eta_{xy} + \eta_{yx}}{4\eta_{yy}}. \qquad (4)$$

Detailed derivations can be found in Supplementary Material [45].

Equation (2) reveals that the $q_x$-term (mediated by $k_y$) deflects $k_x$, which creates an effective gauge field [43], driving asymmetric wave propagation in the x-direction. Under non-Hermitian modulation, this gauge field becomes complex, inducing non-Bloch wave dynamics characterized by directional amplification or attenuation. When encountering boundaries, wave localizations occur, precisely manifesting the NHSE. Similarly, the $q_y$-term mediated by $k_x$ leads to non-Bloch wave dynamics in the y-direction, which, cooperating with the $q_x$-term, generates higher-order NHSE.

As a hallmark of NHSE, the bulk-boundary correspondence is fundamentally broken. To demonstrate this, we truncate the medium along the *x*-direction by hard boundaries, as illustrated in Fig. 1(a). Its dispersion relation is given as (see Supplementary Material [45])

$$\frac{\omega_{w/B}^2}{K} = \eta_{xx}\left(\frac{\pi n}{L}\right)^2 + \eta_{yy}k_y^2 - 4\eta_{xx}q_x^2 k_y^2, n = 0, 1, 2, 3, \ldots \quad (5)$$

Here, $\omega_{w/B}$ denotes eigenfrequencies with boundaries, in contrast to $\omega_{w/o}$ in Eq. (2) representing eigenfrequencies in the bulk. Figure 1(b) visualizes their difference. $\omega_{w/o}$ (red-blue) forms open loops for $k_y \neq 0$, where forward ($k_x > 0$) and backward ($k_x < 0$) waves experience asymmetric amplification/attenuation rates, generating unidirectional energy flow. Upon encountering boundaries, energy becomes localized, collapsing $\omega_{w/B}$ (grey) into straight arcs.

We point out that the inconsistency of dispersion relations with and without boundaries is a smoking-gun signature of NHSE in uniform media, similar to the broken bulk-boundary correspondence in discrete lattices. But differently, due to homogeneity, the bulk dispersion loops for uniform media are not closed, posing a curiosity on the applicability of point-gap topology in such systems. In fact, the literature have suggested modified winding numbers by tracing open spectral loops, which may promise novel topological properties [40,41].

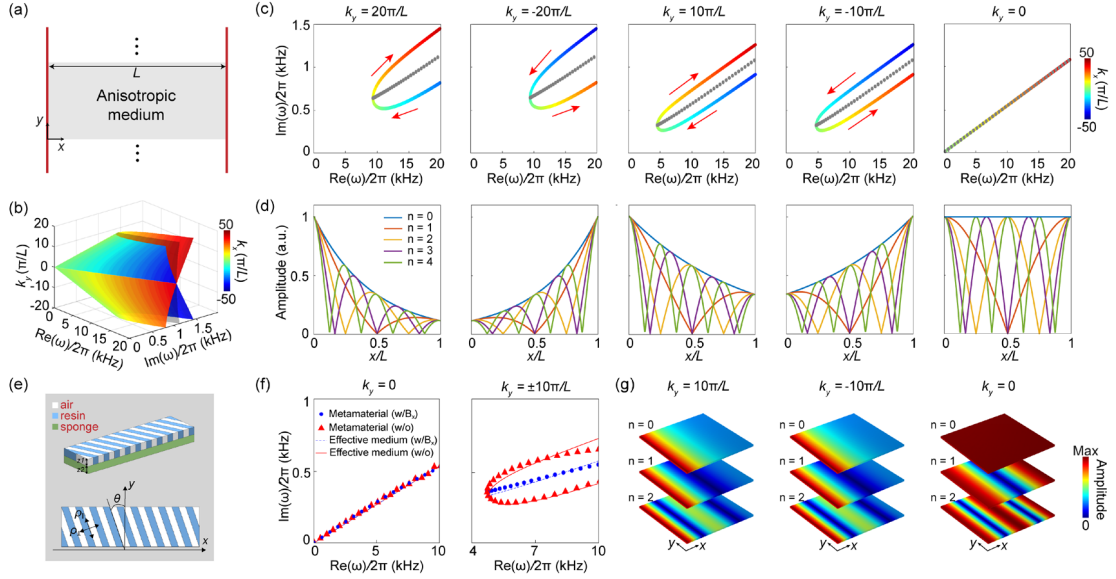

FIG. 2. (a) Sketch of a slice of uniform medium with hard boundaries in the $x$-direction. (b) Comparison of dispersions between an infinite medium ($\omega_\text{w/o}$, red-blue surface) and the finite medium in (a) ($\omega_\text{w/B}$, grey surface). The material parameters are set as $\rho_{xx} = \rho_{yy} = 3.08 - 0.43i$ kg/m$^3$, $\rho_{xy} = \rho_{yx} = 1.12 - 0.27i$ kg/m$^3$, and $K = 222$ kPa. (c) Cross-sections of (b) at selected $k_y$. Arrows trace $k_x$-trajectory from negative to positive. (d) Pressure field amplitudes for the first five $\omega_\text{w/B}$ eigenmodes. (e) An acoustic multilayer metamaterial implementing the uniform medium. Its feasibility is verified by simulated (f) $\omega_\text{w/o}$-$\omega_\text{w/B}$ dispersions and (g) pressure field amplitudes, which closely match those of the uniform medium. Geometric parameters: $L = 200\sqrt{2}$ mm.

The dispersion inconsistency is further quantified for selected cases in Fig. 2(c). Crucially, $k_y$ governs both the trajectory of $\omega_\text{w/o}$ and its winding direction (as $k_x$ sweeps from negative to positive). This is directly aligned with the description of effective gauge field, which is induced by anisotropy but mediated by wave vectors [see Eq. (2)]. This $k_y$-dependence enables flexible control of NHSE. For the finite medium in Fig. 2(a), the pressure field amplitude follows (see Supplementary Material [45]).

$$|p(x,y)| = 2e^{-\text{Im}(2q_x k_y)x} e^{\text{Im}(k_y)y} \left|\cos\left(\frac{\pi n}{L}x\right)\right|. \quad (6)$$

As visualized in Fig. 2(d), $\text{Im}(q_x k_y) > 0$ produces left-boundary localization, while

$\text{Im}(q_x k_y) < 0$ yields right-boundary localization. At $\text{Im}(q_x k_y) = 0$ (where the effective gauge field vanishes), the Bloch condition is restored, featuring extended standing waves. Notably, here $k_y$ is taken as purely real for clarity. With complex $k_y$, non-Bloch wave manipulation can be extended to the $y$-direction, enabling wave funneling along designed paths and localization at targeted boundaries.

To realize the anisotropic uniform medium, we design an acoustic metamaterial consisting of alternating layers of air and acoustic rigid materials (e.g., photosensitive resin, a 3D printable material), as illustrated in Fig. 2(e). This multilayer design exhibits effective anisotropic mass density ($\rho_\parallel$, $\rho_\perp$), with non-Hermiticity introduced by adding a uniform layer of sponge (an acoustic lossy material) at the bottom of the metamaterial. Details of the effective parameter calculations are provided in Supplementary Material [45].

With anisotropic ($\rho_\parallel$, $\rho_\perp$), complex gauge fields $q_x$ and $q_y$ can be generated by rotating the metamaterial, yielding the density tensor

$$\rho = \begin{pmatrix} \rho_\perp \cos^2\theta + \rho_\parallel \sin^2\theta & (\rho_\perp - \rho_\parallel)\cos\theta\sin\theta \\ (\rho_\perp - \rho_\parallel)\cos\theta\sin\theta & \rho_\parallel \cos^2\theta + \rho_\perp \sin^2\theta \end{pmatrix}, \qquad (7)$$

where $\theta$ denotes the rotation angle, which controls the strength of the gauge fields. As an example, we consider $\theta = 45°$, which presents effective material parameters consistent with the uniform medium, as verified by matched dispersions shown in Fig. 2(f). Simulated pressure field amplitudes in the metamaterial at different $k_y$ further confirm the emergence of NHSE [see Fig. 2(g)].

It is important to note that the specific example in Figs. 2(f-g) is used solely to validate the data in Figs. 2(b-d). In fact, NHSE occurs for broad (even random) material parameters, as long as $\text{Im}(q_x k_y) \neq 0$ (see Supplementary Material [45]). This is precisely due to our novel mechanism of generating effective complex gauge fields using anisotropic material parameters, which significantly relaxes non-Hermitian modulations compared to the fine-tuned discrete models. In our design, a simple uniform layer of lossy material is sufficient for the required non-Hermiticity. This makes the practical applications of NHSE highly feasible, especially for on-chip integrations, where the ease of gain/loss control can substantially reduce the fabrication complexity.

NHSE in uniform media allows wave guiding with a very distinctive mechanism. As Eq. (6) indicates, wave amplitude is governed by non-Bloch wave dynamics in the $x$- and $y$-directions, characterized by the amplification/attenuation rates $\text{Im}(q_x k_y)$ and $\text{Im}(k_y)$, respectively. By modulating the effective gauge field and the wave vector, both wave trajectory and localization can be precisely engineered. This guiding principle is termed non-Hermitian wave funneling. Of the two governing parameters, while $q_x$ is determined by material anisotropy, $k_y$ relies on excitation conditions. In the preceding analyses, we focused on the discussions of eigenspectra and eigenmodes with real $k_y$ (for simplicity), yielding complex $\omega_{\text{w/B}}$, as described by Eq. (5). Alternatively, this eigenvalue problem can be reformulated by imposing real $\omega_{\text{w/B}}$, enabling complex $k_y$ as

$$k_y = \pm\sqrt{\frac{\frac{\omega_{\text{w/B}}^2}{K}-\eta_{xx}\left(\frac{\pi n}{L}\right)^2}{\eta_{yy}-4\eta_{xx}q_x^2}}, n = 0, 1, 2, 3, ... \tag{8}$$

We emphasize that both real and complex $k_y$ are eigenvalues of Eq. (5), manifesting the NHSE. Crucially, the complex $k_y$ not only regulates the effective gauge field along the $x$-direction via $\text{Im}(q_x k_y)$, but also contributes to the non-Bloch wave dynamics along the $y$-direction (see more discussions in Supplementary Material [45]). Moreover, complex $k_y$ corresponds to real $\omega_{\text{w/B}}$, making it more suitable for wave control applications where excitations typically operate at real frequencies.

**Experimental realization**

For demonstration, we fabricate an acoustic metamaterial sample as shown in Fig. 3(a). A point source with real excitation frequency is placed at the bottom of the sample to launch upward-propagating waves, corresponding to $\text{Im}(q_x k_y) > 0$. Figure 3(b) presents the experimentally measured wave trajectory (see set-up and measurement details in Supplementary Material [45]). It shows that despite being semi-directional, wave energy from the point source is entirely funneled leftward and eventually localized at the left-boundary. Conversely, when the source is placed at top of the metamaterial, corresponding to downward propagation with $\mathrm{Im}(q_x k_y) < 0$, wave funneling occurs toward the right boundary [Fig. 3(c)]. These observations are well aligned with above theoretical analyses. Notably, the point source excites a wide range of wave vector components, revealing collective funneling behavior. By tuning individual incident angle or excitation location, the funneling path can be customized (see Supplementary Material [45]). Beyond wide-angle operation, the non-Hermitian wave funneling persists under broadband excitation, as demonstrated in Fig. 3(d).

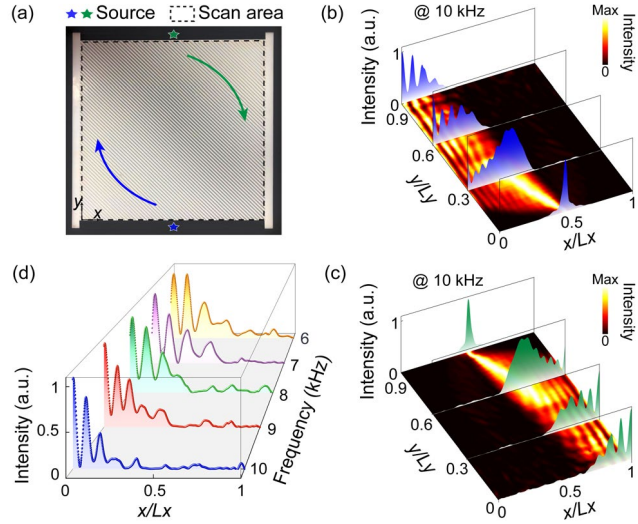

FIG. 3. (a) Fabricated metamaterial sample demonstrating the non-Hermitian wave funneling, where upward-propagating waves from a point source (blue) are guided toward the left-boundary, while their downward counterparts (green) are guided toward the right-boundary. Experimental validation is presented in (b) for leftward and in (c) for rightward funneling. Measured wave intensities are normalized at each *y*-scanning step, with cut-line plots highlighting the intensity evolution along the funneling paths. (d) Cut-line plots for leftward funneling under broadband excitation (see full intensity distributions in Supplementary Material [45]).

By direct generalization, we next demonstrate the realization of higher-order NHSE. Consider a piece of anisotropic medium with hard boundaries in both the *x*- and *y*-directions. As illustrated in Fig. 4(a), the effective gauge fields $q_x$ and $q_y$ jointly manifest, inducing non-Bloch wave localization at corners, known as the corner skin modes, which hallmark the higher-order NHSE. To reveal the NHSE, we numerically calculate the eigenspectra for the finite medium (using identical material parameters to Fig. 2). The results are presented in Fig. 4(b) (left panel). For comparison, spectra without boundaries are projected onto the $\text{Re}(\omega)$-$\text{Im}(\omega)$ plane. The observed spectral inconsistency with versus without boundaries unambiguously confirms NHSE. Further eigenmode analysis in Fig. 4(b) (right panel) verifies wave localization at corners. Owing to inversion symmetry of the density tensor [Eq. (7)], the corner skin modes emerge simultaneously at either upper-left/lower-right corners (along the $\rho_\parallel$ direction, modes #1 and 2) or lower-left/upper-right corners (along the $\rho_\perp$ direction, modes #3 and 4). Notably, as a result of the complex gauge fields, modes #1 and 2 exhibit lower decay rates than modes #3 and 4. This is because acoustic waves preferentially propagate through air layers along the $\rho_\parallel$ direction. Conversely, along the $\rho_\perp$ direction, rigid resin layers block waves and divert them through the sponge layer, increasing the decay rates. Additionally, modes #1 and 2 are closer to the real axis in the complex frequency spectrum, and thus couple more efficiently to real-frequency excitations [44].

Asymmetric decay rate as a key signature of the higher-order NSHE enables non-Hermitian wave funneling to specific corners. To demonstrate this, we fabricate a metamaterial sample with hard boundaries in both the *x*- and *y*-directions [see Fig. 4(c)]. According to the eigenmode analysis, wave energy from a point source will be guided and collected at the upper-left and lower-right corners along the $\rho_\parallel$ direction due to the low decay rate, as indicated by the arrow trajectories. This is indeed confirmed experimentally in Fig. 4(d), revealing a broadband corner funneling. This unique wave control mechanism, arising from the synergy between material anisotropy and non-Hermiticity, offers exceptional flexibility compared to the fine-tuned discrete lattice models. Beyond tolerance to wide incident angles, broadband frequencies, and relaxed material constraints, this mechanism accommodates diverse boundary types and shapes, free from lattice restrictions (see Supplementary Material [45]).

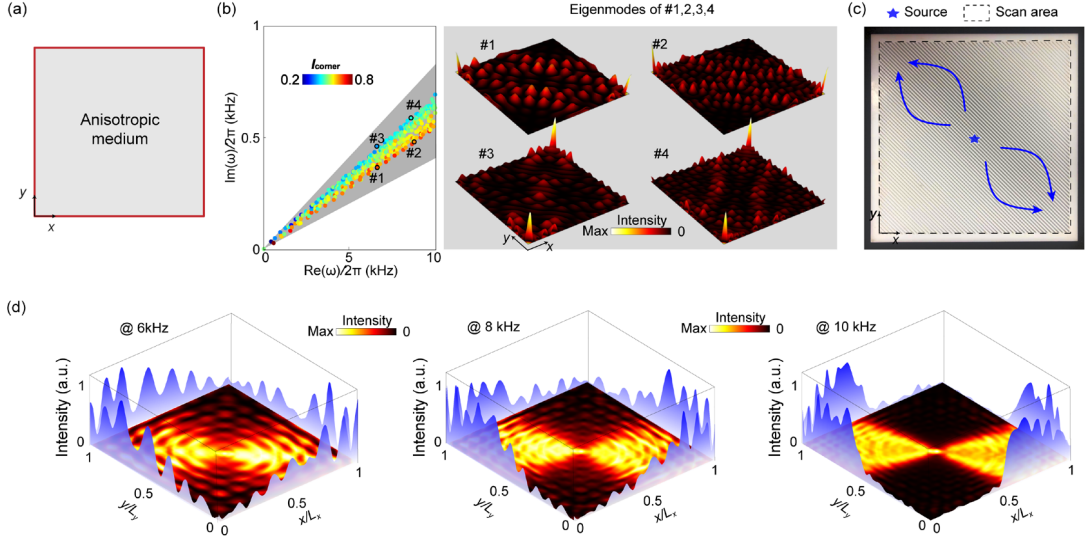

FIG. 4. (a) Illustration of higher-order NHSE in anisotropic medium with hard boundaries in both the *x*- and *y*-directions. Corner skin modes emerge from the combined action of gauge fields $q_x$ and $q_y$. (b) (Left) Eigenspectral inconsistency with (color-coded scatterers) and without (grey shading) boundaries confirms the NHSE. The color indicates the summed intensity of the upper-left and lower-right corners (see calculation details in Supplementary Material [45]). (Right) Intensity distributions for selected eigenmodes, revealing two types of corner skin modes: modes #1/2 (along the $\rho_\parallel$ direction) with low decay rates and modes #3/4 (along the $\rho_\perp$ direction) with high decay rates. Such asymmetric decay rates enable wave funneling to the upper-left and lower-right corners. (c) Fabricated metamaterial sample demonstrating the corner funneling. (d) Experimentally measured funneling paths under point-source excitation at 6, 8, and 10 kHz. The intensity is normalized along the radial direction from the source.

**Conclusions**

In conclusion, we have proposed a novel mechanism for NHSE based on anisotropic media with uniform non-Hermitian modulations. This approach goes beyond previous reliance on fine-tuned discrete models and enables a unique non-Hermitian wave funneling, which can guide wave energy along designed paths toward targeted boundaries. As experimental demonstrations, we have implemented wave funneling using an acoustic multilayer metamaterial, showcasing its broadband and wide-angle performance. The collaboration between material anisotropy and non-Hermiticity introduces a new paradigm for NHSE, not only broadening our understanding of this unique phenomenon, but also providing a novel and practical mechanism for non-Hermitian wave control. Crucially, by eliminating the need for fine-tuned non-Hermitian modulations, this mechanism significantly reduces the fabrication complexity and enhances the feasibility for real-world applications such as energy harvesting and on-chip wave funneling.

**Acknowledgments**

We acknowledge support from the National Key R&D Program of China (Grants No. 2023YFA1407700 and No. 2023YFA1406904), the National Natural Science Foundation of China (Grant No. 12222407) and the Key R&D Program of Jiangsu Province (Grant No. BK20232015). Y.T. thanks the supports from the China Postdoctoral Science Foundation (Grant No. 2023M731614 and Grant No. 2023T160298), the Jiangsu Funding Program for Excellent Postdoctoral Talent (Grants No. 2023ZB114).